\documentclass{iopart}
\usepackage[latin1]{inputenc}
\usepackage{amsfonts}
\usepackage{amssymb}
\usepackage{amssymb,amsfonts,latexsym,fancyhdr,graphicx}

\begin{document}

\title{\small Tomographic approach to the violation of Bell's inequalities
for quantum states of two qutrits}

\author{Bruno Leggio$^{1}$, Vladimir I Man'ko$^{2}$,
Margarita A Man'ko$^{2}$, and Antonino Messina$^{1}$}
\address{$^{1}$Dipartimento di Scienze Fisiche ed Astronomiche
dell'Universit\`{a} di Palermo, Via Archirafi 36, 90123 Palermo,
Italy\\$^{2}$P. N. Lebedev Physical Institute, Russian Academy of
Sciences, Leninskii Prospect 53, Moscow 119991, Russia}

\newcommand{\ket}[1]{\displaystyle{|#1\rangle}}
\newcommand{\bra}[1]{\displaystyle{\langle #1|}}

\begin{abstract}
The tomographic method is employed to investigate the presence of
quantum correlations in two classes of parameter-dependent states
of two qutrits. The violation of some Bell's inequalities in a wide
domain of the parameter space is shown. A comparison between the
tomographic approach and a recent method elaborated by Wu, Poulsen, and
M{\o}lmer shows the better adequacy of the former method with
respect to the latter one.
\end{abstract}

\section{Introduction}

In quantum mechanics, the correlation of physical observables can obey
some constraints expressed in terms of inequalities. For example, the
correlations of position and momentum of a particle provide the
lower bound for the product of the uncertainties of these two quantities as given by the
Schr\"odinger--Robertson uncertainty relation~\cite{Schr30,Robertson30}.

For discrete variables like spins, the quantum correlations for entangled
states, e.g., of two qubits, also provide some inequalities known as
Cirel'son's bound~\cite{Cirelson} $2\sqrt2$, which differs for similar
CHSH--Bell bound 2~\cite{be,CHSH} associated with classical correlations
typical for separable states of two qubits.

In principle, the existence of bounds of this kind can be checked experimentally.
The violation of Bell's inequalities, that is the violation of
the bound of Bell's number 2, was indeed proved experimentally (see, for example,
\cite{Aspect}).

Statistical characteristics as means, dispersions, correlations for
quantum observables can easily be formulated in the probability representation
of Quantum Mechanics (see, recent review~\cite{im}). This representation for
spins was introduced in \cite{DodPL,OlgaJETP}. The quantum states in the probability
representation are described by fair measurable probability distributions
(called tomograms), which contain complete information on density operators
of the states. In view of the possibility to use the standard formulae of
classical probability theory in the probability representation
for quantum-mechanical calculations with tomograms, there might exist some advantages to better understand the role of quantum correlations and of the corresponding quantum bounds for separable
and entangled states, if one uses the tomographic-probability approach.

Our goal in this paper is to investigate the existence of nonclassical correlations
in selected two-qutrits states through Bell's number $B$ using the tomographic approach.

\section{A short review of the quantum tomographic method}

Quantum states are usually described in the canonical quantum picture by vectors in a complex Hilbert space or by density operators, describing pure and mixed states respectively. Together with the usual Schr\"odinger and Heisenberg picture there exist another possible picture of quantum mechanics called Tomographic picture \cite{im}, where a quantum state is described by a quantity called \emph{tomogram}. By definition a tomogram is a positive measurable probability distribution function of random variables. It has been shown that all the physical properties of a quantum system in a given state can be studied in terms of its associated tomogram, which then contains the same amount of information as the density matrix of the state.\\
Although the approach based on tomograms and the one based on density matrices are equivalent, it is worth noting two main conceptual differences between these two methods: first of all in the tomographic picture one works only with classical probability distribution related to relative frequencies of measures. Another difference lies in the fact that the construction of a tomogram requires additional information about some selected observables in a given experimental setup.\\
In the tomographic approach the observables are no longer described by operators: they are now associated with functions. The quantization scheme is then based on a star product of functions \cite{OVM}. \\
This probability representation of Quantum Mechanics provides us with a useful way for analyzing the existence of non local correlations in quantum states of qudits by the use of the so called \emph{qubit portrait} \cite{lm}. This method belongs to the class of entanglement-detecting methods based on linear maps. A qubit portrait is indeed a linear map defined in the tomographic description of quantum states. Such a map constructs a fictious qubit-state associated with the starting qudit-state. This construction is needed because Bell's inequality are defined only for systems described by dichotomic variables.\\
In the following chapters this method will be applied to check Bell's inequalities in two classes of quantum states.

\section{Tomographic analysis of entangled states of two qutrits}

We know from Bell's theorem~\cite{be} that the existence of nonclassical correlations
is given by a violation of the inequality
\begin{equation}B\leq2.\end{equation}
Using the qubit-portrait method it has been shown \cite{cm} that the Bell's number for
the maximally entangled state of two qutrits
\begin{equation} \ket{\psi}=\frac{1}{\sqrt{3}}\{\ket {11}+\ket{00}+\ket{-1-1} \} \end{equation}
can take values greater than 2, for example being $B=1+\sqrt{2}$.

In this letter we aim at extending such a result by applying the tomographic method to the following two classes of entangled states
\begin{eqnarray}\ket{\psi_{\varphi}}&=&\frac{1}{\sqrt{3}}\{\ket {1-1}
+\ket{0 0}+e^{-i\varphi}\ket{-11} \},\\
\ket{\psi_{\varphi ,a}}&=&\frac{1}{\sqrt{2+a^{2}}}\{\ket {1-1}+a\ket{0 0}
+e^{-i\varphi}\ket{-11} \},\end{eqnarray}
where the real parameters vary in the ranges $\varphi \in [0,2\pi]$ and $a\in [0,1]$.

\subsection{$\ket{\psi_{\varphi}}$ state}
In order to apply the tomographic method to a quantum state,
first of all we need to specify the associated density matrix.
For our state, this matrix, expressed in the standard basis $\{\ket {1-1},\ket {1 0}, \ket {11}, \ket {0-1}, \ket {0 0}, \ket {0 1}, \ket {-1-1}, \ket {-1 0}, \ket {-1 1}\}$, is denoted by $\rho_{\varphi}$ and has the following explicit form:\\
\begin{equation}
\rho_{\varphi}= \frac{1}{3}\left(\begin{array}{cccccccccc} 1 & 0 & 0 & 0 & 1 & 0 & 0 & 0 & e^{i\varphi}\\
0 & 0 & 0 & 0 & 0 & 0 & 0 & 0 & 0 \\ 0 & 0 & 0 & 0 & 0 & 0 & 0 & 0 & 0 \\ 0 & 0 & 0 & 0 & 0 & 0 & 0 & 0 & 0 \\ 1 & 0 & 0 & 0 & 1 & 0 & 0 & 0 & e^{i\varphi}\\
0 & 0 & 0 & 0 & 0 & 0 & 0 & 0 & 0 \\ 0 & 0 & 0 & 0 & 0 & 0 & 0 & 0 & 0 \\ 0 & 0 & 0 & 0 & 0 & 0 & 0 & 0 & 0 \\ e^{-i\varphi} & 0 & 0 & 0 & e^{-i\varphi} & 0 & 0 & 0 & 1
\end{array}\right).\\
\end{equation}
To obtain the tomogram associated with $\rho_{\varphi}$~\cite{mm}, we must evaluate the nine diagonal elements of the matrix $\tilde{\rho_{\varphi}}$ expressing the density operator $\rho_{\varphi}$ after a generic rotation of the cartesian reference frame, paramatrized in terms of Euler angles. To this end we start by reminding that the matrix rotating the density operator of a single qutrit may be cast in the form\\
\begin{eqnarray}
&&D(\theta,\phi,\chi)=
 \left(
 \begin{array}{cccc}
 \frac{1}{2}e^{-i(\phi+\chi)}(1+\cos \theta) &
 -\frac{1}{\sqrt{2}}e^{-i\phi}\sin \theta & \frac{1}{2}e^{-i(\phi-\chi)}(1-\cos \theta)
 \\ \frac{1}{\sqrt{2}}e^{-i\chi}\sin \theta & \cos \theta & -
 \frac{1}{\sqrt{2}}e^{i\chi}\sin \theta \\ \frac{1}{2}e^{i(\phi-\chi)}
 (1-\cos \theta) & \frac{1}{\sqrt{2}}e^{i\phi}\sin \theta & \frac{1}{2}e^{i(\phi+\chi)}(1+\cos \theta)
\end{array}
\right)\nonumber\\
&&
\end{eqnarray}
where $\theta$, $\phi$, and $\chi$ denote the three Euler angles \cite{sak}.
Hence the rotation matrix to be used to rotate our two-qutrits state $\rho_{\varphi}$ is
\begin{eqnarray} D_{pair}(\theta_{1},\phi_{1},\chi_{1};
\theta_{2},\phi_{2},\chi_{2})=D(\theta_{1},\phi_{1},\chi_{1})\otimes
D(\theta_{2},\phi_{2},\chi_{2}).\end{eqnarray}
where the pedix 1 (2) refers to the qutrit $\vec{S_{1}}$ ($\vec{S_{2}}$).
Evaluating all the nine diagonal elements of the matrix $\tilde{\rho_{\varphi}}=D_{pair}^{\dag}\rho_{\varphi} D_{pair}$ one obtains the following nine-component probability vector
\begin{equation}
\overrightarrow{W}\equiv\left(W_{1},W_{2},W_{3},W_{4},W_{5},W_{6},W_{7},W_{8},W_{9}\right).
\end{equation}
called tomogram of the state $\rho_{\varphi}$. We notice that $\overrightarrow{W}$ is independent on $\chi$ due to the specific form of $\rho_{\varphi}$.\\
To characterize a state through this tomogram amounts at describing it using the density-matrix formalism.
Nevertheless, if we want to check the Bell's inequality, it is easier to start from the tomogram of the state since it allows a systematic construction of a new representation mimising a qubit tomogram called qubit portrait ~\cite{lm}. In practice starting from our nine-component probability vector $\overrightarrow{W}$ we define the following new four-component tomogram
\begin{equation} \overrightarrow{\omega}\equiv(\omega_{1},\omega_{2},\omega_{3},\omega_{4}), \end{equation}
where in accordance with \cite{cm} we put
\numparts
\begin{eqnarray}
\omega_{1}=W_{1}, \label{portr1}\\
\omega_{2}=W_{2}+W_{3},\\
\label{portr2}
\omega_{3}=W_{4}+W_{7},\\
\label{portr3}
\omega_{4}=W_{5}+W_{6}+W_{8}+W_{9}.
\label{portr4}
\end{eqnarray}
\endnumparts
$\overrightarrow{\omega}$ represents the tomogram of a quantum state of a fictious two-qubit system. This mapping procedure qutrit $\longrightarrow$ qubit provides the way to construct the Bell's number of our original state $\rho_{\varphi}$.\\
From an experimental point of view to check the Bell's number of a two qubit system we must fix for each qubit in a generic way two orientations of our measuring apparatus. In this way we perform two measurements for each qubit. We denote by $a=(\theta_{1a},\phi_{1a})$ and $b=(\theta_{1b},\phi_{1b})$ ($c=(\theta_{2c},\phi_{2c})$ and $d=(\theta_{2d},\phi_{2d})$) the orientations of the apparatus corresponding to the measurements performed on qubit 1 (2).
To construct the expression of $B$ for our state $\rho_{\varphi}$ it is useful to introduce the $4\times 4$ matrix
\begin{equation}
T=\left(\begin{array}{cccc} \overrightarrow{\omega_{ac}} &
\overrightarrow{\omega_{ad}} & \overrightarrow{\omega_{bc}} & \overrightarrow{\omega_{bd}}
\end{array}\right),
\end{equation}
where $\overrightarrow{\omega_{ij}}$ is a $4\times 1$ column vector obtained from the vector
$\overrightarrow{\omega}$ by formally assigning to the angle variables
$\theta_{1}$, $\phi_{1}$, $\theta_{2}$, and $\phi_{2}$ the generic but fixed values $\theta_{1i}$, $\phi_{1i}$, $\theta_{2j}$, and $\phi_{2j}$ ($i=a,b ; j=c,d$).
The Bell's number of our original state $\rho_{\varphi}$ can then be expressed as
\begin{equation}
B=|\mbox{Tr}\,(IT)|,
\end{equation}
where $I$ is the polarization matrix defined as
\begin{equation}
I=\left(\begin{array}{ccccc} 1 & -1 & -1 & 1\\
1 & -1 & -1 & 1 \\ 1 & -1 & -1 & 1 \\ -1 & 1 & 1 & -1
\end{array}\right).
\end{equation}
It is worthnoting that $B$ is a function of the eight angles $\theta_{1a}$, $\theta_{1b}$, $\theta_{2c}$,
$\theta_{2d}$, $\phi_{1a}$, $\phi_{1b}$, $\phi_{2c}$, and $\phi_{2d}$ so that we can look for a violation of Bell's inequality by maximizing $B$
with respect to such eight parameters.
The result of such a maximization is still a function of the state parameter $\varphi$.
Due to the mathematical difficulties of such a procedure, we have to resort to a numerical maximization. The function of $\varphi$ obtained
in this way is shown in Fig.~1.
\begin{figure}[h]
\begin{center}
\includegraphics[width=200pt]{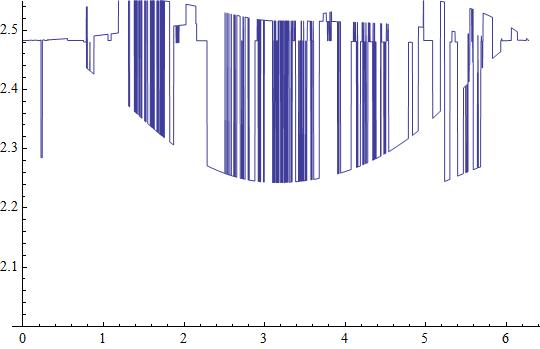}
\end{center}
\caption{Bell's number of the $\ket{\psi_{\varphi}}$ state, where
$\varphi$ is on the abscissa the numerical maximization, which gives the Bell's number,
is on the ordinate.}
\end{figure}\\
It is obvious that the Bell's inequality is violated
in the whole range of variability of the parameter $\varphi$.
Hence in our state nonclassical correlations are present no matter which value we
give to the phase parameter.

\subsection{$\ket{\psi_{\varphi,a}}$ state }

Next, we investigate the existence of nonclassical correlations in the two-parameter normalized state
\begin{equation}\ket{\psi_{\varphi ,a}}=\frac{1}{\sqrt{2+a^{2}}}\left\{\ket {1-1}+a\ket{0 0}+
e^{-i\varphi}\ket{-11}\right\},\end{equation}
where $a \in [0,1]$. We follow the same method used in the previous paragraph immediately representing $\rho_{\varphi,a}=\ket{\psi_{\varphi ,a}}\langle \psi_{\varphi ,a}|$ as follows:

\begin{eqnarray}
\rho_{\varphi,a}=
\frac{1}{2+a^{2}}\left(\begin{array}{cccccccccc} 1 & 0 & 0 & 0 & a & 0 & 0 & 0 & e^{i\varphi}\\
0 & 0 & 0 & 0 & 0 & 0 & 0 & 0 & 0 \\ 0 & 0 & 0 & 0 & 0 & 0 & 0 & 0 & 0 \\
0 & 0 & 0 & 0 & 0 & 0 & 0 & 0 & 0 \\ a & 0 & 0 & 0 & a^{2} & 0 & 0 & 0 & ae^{i\varphi}\\
0 & 0 & 0 & 0 & 0 & 0 & 0 & 0 & 0 \\ 0 & 0 & 0 & 0 & 0 & 0 & 0 & 0 & 0 \\
 0 & 0 & 0 & 0 & 0 & 0 & 0 & 0 & 0 \\ e^{-i\varphi} & 0 & 0 & 0 & ae^{-i\varphi} & 0 & 0 & 0 & 1
\end{array}\right).
\end{eqnarray}\\
Once again we have to rotate this density matrix exploiting $D_{pair}(\theta_{1},\phi_{1},\chi_{1};
\theta_{2},\phi_{2},\chi_{2})$ and then taking the diagonal elements of this rotated matrix $\tilde{\rho_{\varphi,a}}=D_{pair}^{\dag}\rho_{\varphi,a} D_{pair}$ in order to construct
the tomogram $\overrightarrow{\mathcal W}$ of our state $\rho_{\varphi,a}$. Proceeding as in the previous section, we consider a qubit-portrait
of our qutrit state in the same way as we did for the $\ket{\psi_{\varphi}}$ state.
Using Equations~(10), we obtain
\begin{eqnarray}
\mathit w_{1}=\mathcal W_{1},\nonumber\\
\mathit w_{2}=\mathcal W_{2}+\mathcal W_{3},\nonumber\\
\mathit w_{3}=\mathcal W_{4}+\mathcal W_{7},\nonumber\\
\mathit w_{4}=\mathcal W_{5}+\mathcal W_{6}+\mathcal W_{8}+\mathcal W_{9}.\nonumber
\end{eqnarray}
Proceeding with the method exposed above, after a numerical maximization on
the set $\left\{\theta_{1a},\theta_{1b},\theta_{2c},\theta_{2d},\phi_{1a},
\phi_{1b},\phi_{2c},\phi_{2d}\right\}$, we obtain a function of two parameters
$\varphi$ and $a$.

Once again an analytical approach is quite complicated and hence after a numerical treatment we plot this function of both parameters getting the surface presented in Fig.~2.
\begin{figure}[h]
\begin{center}
\includegraphics[width=200pt]{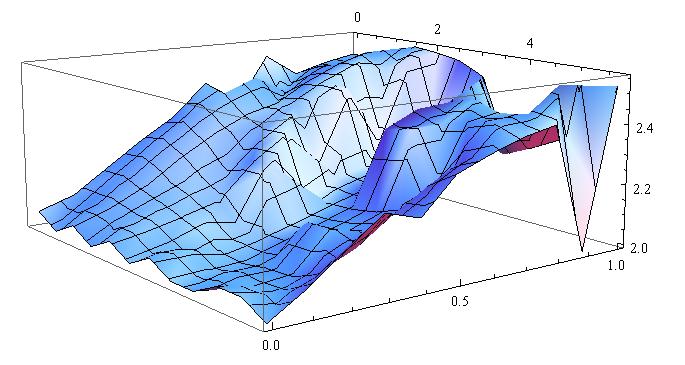}
\end{center}
\caption{Bell's number for the $\ket{\psi_{\varphi,a}}$ state plotted against
the two parameters $\varphi$ and $a$.}
\end{figure}\\
The most striking feature of this figure is the presence of domains in the $(\varphi , a)$-plane  where a sudden change
in the value of Bell's number occurs. In particular, we notice the existence of a domain
around the point $(\varphi_0,a_0)\sim({11\pi}/{6},\,0.9)$ where the Bell's number goes down
to a value around 2.
In order to see this behaviour in more detail we report in Fig.3 the Bell's number of our state
against $a$ keeping $\varphi$ fixed at ${11\pi}/{6}$.
\begin{figure}[h]
\begin{center}
\includegraphics[width=200pt]{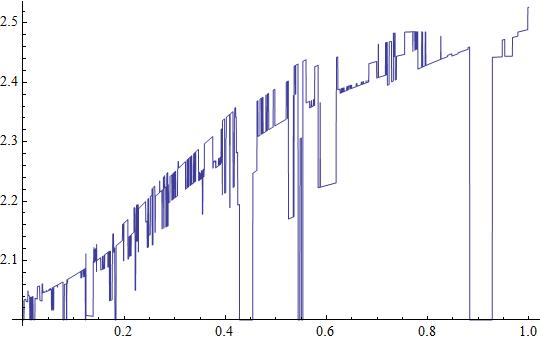}
\end{center}
\caption{Bell's number for the $\ket{\psi_{\varphi,a}}$ state plotted against
the parameter $a$ when $\varphi$ is fixed at the value $ \varphi_{0}={11\pi}/{6}$.}
\end{figure}\\
This graph clearly shows the existence of domains in which the Bell's number
of our state crosses the critical value 2. For example keeping $a$ and $\varphi$ fixed at values $(\varphi_0,a_0)\equiv({11\pi}/{6},\,0.9)$ it is possible to proceed with an exact maximization of the Bell's number which gives the result $B_{0}=1.99999999976005$. This means that the state satisfies the Bell's
inequality in that point.

This anomalous behaviour may be due to many effects.

First of all, it could just be an effect of our numerical approach to the problem
or it could mean that our choice of the qubit portrait is no longer correct in this domain
of the parameter space, and we should test the value of Bell's number for every other
possible qubit portraits. However, any other choice in our qubit portrait we tested
gives a similar result. If this behaviour is not due to numerical maximization problems
or to unfortunate choices of qubit maps, this could mean that our state undergoes a transition
from the condition in which nonclassical correlations are present to a condition in which
the only correlations present in the state are of a classical kind. If this is the case then there exists a critical value of the parameter $a$ at which this transition takes place.

It is of interest to observe that the time-dependent state
\begin{equation}\label{psit}
\ket{\psi_{t}}=\frac{1}{\sqrt{2+\cos(\omega t)^{2}}}\left\{
\ket {1-1}+\cos(\omega t)\ket{0 0}+e^{-i{11\pi}/{6}}\ket{-11}\right \}.
\end{equation}
is such to span progressively and periodically all the states in the class $\{ \ket{\psi_{11\pi \setminus6,a}}, a \in [0,1] \}$. As a consequence it evolves from the initial condition
\begin{equation}\label{psi0}
\ket{\psi_{0}}=\frac{1}{\sqrt{3}}\left\{\ket {1-1}+
\ket{00}+e^{-i{11\pi}/{6}}\ket{-11}\right \},\end{equation}
in such a way that when $\omega t \simeq \arccos(a_{0})$ the Bell's number is less than 2. To exploit the properties exhibited by $\ket{\varphi , a}$ thus one might wonder whether an Hamiltonian $H(t)$ exists such that $\ket{\psi (t)}$ satisfies the corresponding Schr\"{o}dinger equation
\begin{equation}\label{schr}
i\frac{\partial \ket{\psi (t)}}{\partial t}=H(t)\ket{\psi (t)}
\end{equation}
To this end consider the Hamiltonian
\begin{equation}\label{ham}
H(t)=\Omega_{0}(S_{1}^{z}+S_{2}^{z})+\lambda(t)\big[iS_{1}^{z}S_{2}^{z}(S_{1}^{+}S_{2}^{-}+e^{-\frac{i11\pi}{6}}S_{1}^{-}S_{2}^{+})-i(e^{\frac{i11\pi}{6}}S_{1}^{+}S_{2}^{-}+S_{1}^{-}S_{2}^{+})S_{1}^{z}S_{2}^{z}\big].
\end{equation}
When the coupling constant $\lambda(t)$ has the form
\begin{equation}
\lambda(t)=-\frac{\omega\sin(\omega t)}{5+\cos(2\omega t)}
\end{equation}
and $\Omega_{0}$ is an arbitrarily assigned energy.\\
It is easy to verify that the time evolution operator $U(t)$ associated with $H(t)$ may be written down as
\begin{equation}
U(t)=e^{-i\Omega_{0}(S_{1}^{z}+S_{2}^{z})t}\exp\Big(i\Lambda(t)\big[iS_{1}^{z}S_{2}^{z}(S_{1}^{+}S_{2}^{-}+e^{-\frac{i11\pi}{6}}S_{1}^{-}S_{2}^{+})-i(e^{\frac{i11\pi}{6}}S_{1}^{+}S_{2}^{-}+S_{1}^{-}S_{2}^{+})S_{1}^{z}S_{2}^{z}\big]\Big)
\end{equation}
where $\Lambda(t)=\int_{0}^{t}\lambda(x)dx$ and that
\begin{equation}
U(t)\ket{\psi(0)}=\ket{\psi(t)}
\end{equation}
with $\ket{\psi(0)}$ and $\ket{\psi(t)}$ given by (17) and (16) respectively.\\
It is not difficult to convince oneself that starting from (19) other examples of Hamiltonians such that $\ket{\psi(t)}$ satisfies eq. (18) may be constructed.\\ \\

Recently a new method has been proposed to investigate the presence of quantum correlations
in a given quantum state~\cite{mo}. This approach suggests to use the non-negative quantity
$Q(A,B)=S(A,B)-I_{\max}(A,B)$ to characterize the existence of nonclassical correlations
in the state of a quantum bipartite system. Here $A$ and $B$ are meant to be the two subsystems,
$S(A,B)$ is the quantum mutual information and $I_{\max}(A,B)$ is the maximum classical mutual information
over all possible choices of local measurements. The interesting property is that
the function $Q(A,B)$ vanishes only for states wherein the only correlations are classical.
Nevertheless, from a practical point of view, the maximization involved in the evaluation
of $I_{\max}(A,B)$ is often quite complicated. Indeed, classical mutual information $I(A,B)$,
depending strongly on the choice of measurement strategies, should be maximized over
every possible POVM. One route toward the resolution of this problem is to confine
to projective measures only; this, in turn, allows the evaluation of the quantity
$Q^{p}(A,B)=S(A,B)-I^{p}_{\max}(A,B)$ which is meant to surrogate $Q(A,B)$.
Since, indeed, the set of projective measures is just a subset of all POVM, we must have
\begin{equation}Q(A,B)\leq Q^{p}(A,B).\end{equation}
Whether $Q^{p}(A,B)$ may help to extract definitive replies on the
existence of quantum correlations in a bipartite system in a given
quantum state is still an open problem. We note indeed that when
the function $Q^{p}(A,B)$ is positive, we are not able, in general,
to give a definitive answer on the presence of quantum
correlations since, in this case, we do not know anything about the
positivity of $Q(A,B)$. Only when $Q^{p}(A,B)$ vanishes, we can
state with certainty that $Q(A,B)=0$ which, in turn, means the
existence of only classical correlations.

When applied to the
$\ket{\psi_{\varphi,a}}$ state we get $Q^{p}(A,B)$ depending only on $a$ and this result is shown in Fig.~4. Since the class $\ket{\psi_{\varphi}}$ coincides with $\ket{\psi_{\varphi,1}}$, we immediately see that $Q^{p}(A,B)$ in this latter case assumes a constant and positive value deducible from Fig.4.
\begin{figure}[h]
\begin{center}
\includegraphics[width=200pt]{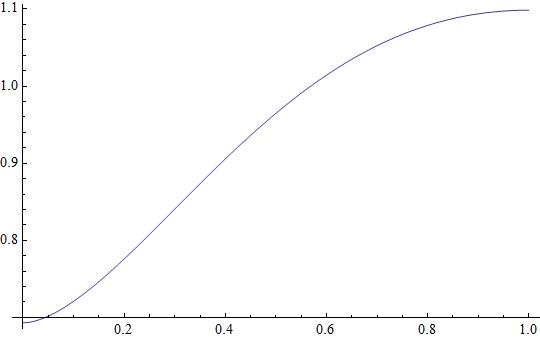}
\end{center}
\caption{Function $Q^{p}(A,B)$ for the $\ket{\psi_{\varphi,a}}$ state plotted against the parameter $a$.}
\end{figure}\\
Hence we see that, for the two-parameter state we are considering, the $Q^{p}(A,B)$ function
cannot reproduce the result we obtained with our tomographic method when the parameter
$a$ crosses its critical value around 0.9. In the case we are considering, the method reported
in \cite{mo} is not able to give any useful answer concerning the presence of quantum correlations.

On the contrary, the tomographic method we applied is a more powerful tool
of analysis for the state we studied. Indeed, the tomographic approach allows us
to give definitive replies about the existence of quantum correlations in a
wide range of the parameter domain where the Bell's number is greater than $2$.

\section{Conclusions}
We have examined a class of one-parameter two qutrits state using the tomographic approach showing without any doubt the existence of non-classical correlations in all the domain of the parameter.\\
We have also studied a class of two-parameter states of two qutrits finding almost everywhere in the domain of the two parameters non-classical correlations. We find in this class also examples of states which might be compatible with the Bell's inequalities. This fact suggests the possibility of a passage from states exibiting non-classical correlations to states dominated by classical correlations only at the transition point $\sim(\frac{11\pi}{6} , 0.9)$ when $a$ is varied and $\varphi$ is kept at its value $\frac{11\pi}{6}$.\\
In view of such a property we have demonstrated that the sub-class $\{ \ket{\psi_{11\pi \setminus6,a}}, a \in [0,1] \}$ may be dynamically generated and the relevant time dependent Hamiltonian accomplishing this target is explicitly given.\\
The comparison with another available
method~\cite{mo} for studying nonclassical correlations of two-qutrit system shows
that the tomographic approach provides a tool to detect some peculiarities of the
correlations which are not detected by using mutual information of bipartite system.
The results of this paper can be extended to higher spins.

\section*{Acknowledgments}
V.I.M and M.A.M thank the Russian Foundation for Basic Research for partial support
under Projects Nos.~07-02-00598 and 09-02-00142 and the University of Palermo (Italy)
for kind hospitality.

\end{document}